\documentclass[11pt]{article}
\usepackage{amsmath,amssymb}
\usepackage{graphicx}
 \allowdisplaybreaks
 \newcommand{\be}{\begin{equation}}
 \newcommand{\ee}{\end{equation}}
 \newcommand{\bea}{\begin{eqnarray}}
 \newcommand{\eea}{\end{eqnarray}}
 \newcommand{\nn}{\nonumber}
 \newcommand{\td}{\tilde}
 
 \newcommand{\pd}{\partial}
 \newcommand{\one}{{\bf 1}}

 \newcommand{\cO}{{\cal O}}

 \newcommand{\bpsi}{\bar\psi}
 
\long\def\symbolfootnote[#1]#2{\begingroup%
\def\thefootnote{\fnsymbol{footnote}}\footnote[#1]{#2}\endgroup}
\newcommand{\aei}{\it Max Planck Institute for Gravitational Physics
(Albert Einstein Institute)\\ Am M\"uhlenberg 1, 14476 Golm,
Germany}

\newcommand{\auth}{Jianwei Mei}

\begin{document}

%\begin{flushright}
%\hfill{
%MIFP-xxx}\\
%\hfill{
%\bf hep-th/yymmnnn}
%\end{flushright}

\begin{center}

~\vspace{20pt}

\centerline{\Large\bf Towards A Possible Fluid Flow Underlying the
Kerr Spacetime}

\vspace{25pt}

%\auth \symbolfootnote[1]{footnote here...}

\auth

%\vspace{10pt}{\ihep}

%\vspace{10pt}{\tamu}

\vspace{10pt}{\aei}

\vspace{2cm}

\underline{ABSTRACT}

\end{center}

Based on the idea of emergent spacetime, we consider the
possibility that the material underlying our spacetime is modelled
by a fluid. We are particularly interested in possible connections
between the geometrical properties of the emergent spacetime and
the properties of the underlying fluid. We find some partial
results that support this possibility. By using the Kerr spacetime
as an example, we construct from the Riemann curvature tensor a
vector field, which behaves just like the speculated fluid flow.

%%%%%%%%%%%%%%%%%%%%%%%%%%%%%%%
 \newpage
% \setcounter{footnote}{0}
% \setcounter{page}{1}
%%%%%%%%%%%%%%%%%%%%%%%%%%%%%%%

\section{Introduction}

By far the geometrical properties are the only thing that we know
about our spacetime. As a physical entity, however, the space must
have more interesting substructures. But the substructures of the
space may not become obvious until the Planck scale, so it will be
extremely difficult to obtain non-geometrical information about
the spacetime.

In view of this difficulty, we firstly ask a much easier question:
what could be our best guess about the materials underlying the
space? Although an answer to this question does not solve the real
physics problem, it may invite ideas that can help solve the real
problem in the end. Here we want to consider the possibility that
the material underlying our space is modelled by a fluid. Hints
toward such a possibility come from at least two different
directions.

The first is related to a branch of study in the Gauge/Gravity
duality (see, e.g. \cite{aharony.gubser.maldacena.ooguri.oz99,
kovtun.son.starinets04, hartnoll.herzog.horowitz08,
guica.hartman.song.strominger08, herzog.kovtun.son,
liu.mcgreevy.vegh09}). As a general feature, the duality always
involves a gravitational theory in the bulk and a gauge field
theory on the boundary. The gauge field theory describes a matter
system that can often be modelled by a fluid in some appropriate
limit \cite{policastro.son.starinets01,
policastro.son.starinets02,
bhattacharyya.ehubeny.minwalla.rangamani08,
bredberg.keeler.lysov.strominger11,
compere.mcfadden.skenderis.taylor}. It is hoped that the
Gauge/Gravity duality can lead to a consistent formulation of the
quantum gravity theory. If such a hope does become true, it is
then quite probable that the material underlying our spacetime is
related to the gauge theory in the Gauge/Gravity dual, and thus
may behave like a fluid in the appropriate limit.

The second is related to the existence of classical mechanical
analogs of relativistic effects \cite{musienko.manevich04}. Here
the key observation is that in some condensed matter system,
certain excitations of the system only ``see" a Minkowski
spacetime. (A simple example can be found with a chain of
identical harmonic oscillators.) If relativistic-like effects can
arise from a real condensed matter system, then it is possible
that Special Relativity or even General Relativity can also emerge
from some more sophisticated condensed matter systems
\cite{barcelo.liberati.visser05}. One can imagine that all
currently known matter fields are excitations of the material
underlying the space, and our measured spacetime is {\it emergent}
in the sense that it is purely determined by the equations obeyed
by the matter fields.\footnote{Note our meaning of the {\it
emergent spacetime} is rather different from that in
\cite{seiberg06}.}

Of course, the above arguments are merely speculations. So it is
desirable if one can find more concrete evidence to support (or to
reject) the fluid picture. In this note, we want to show that such
a possible evidence dose exist. In \cite{katanaev.volovich92},
defects in solids are related to geometrical quantities such as
the Riemann curvature tensor and the torsion tensor. Here we want
to turn the logic around and ask if one can obtain some
information about the underlying material, by using the
geometrical properties of the emergent spacetime. We will show
that it is possible to construct from the Riemann curvature tensor
a vector field, which behaves just like the speculated fluid flow
underlying our spacetime.

Lets firstly explain what we expect for the proposed fluid. We
assume that the fluid is similar to the ones that we know in the
emergent spacetime. Most importantly, we assume that the fluid is
also characterized by a density $\rho$ and a four velocity
$u^\mu$, satisfying $g_{\mu\nu}u^\mu u^\nu=-1$, where $g_{\mu\nu}$
is the metric of the emergent spacetime. When the emergent
spacetime is flat, we expect the underlying fluid to be static (in
one particular coordinate system) and everywhere uniform. The
presences of matter fields in the emergent spacetime should
correspond to non-trivial disturbances in the underlying fluid.
Especially, when the matter fields have a net angular momentum,
like in the case of a rotating star or a Kerr black hole, there
might be a corresponding global current in the underlying fluid.
In this case, the fluid is still static and uniform at the spatial
infinity, but will have a circulation motion near the center. For
all these features, it is enough to consider only stationary
spacetimes, which are characterized by the presence of a time-like
Killing vector.

Our main purpose of this note is to show that all the above
expectations are fulfilled by the vector field that we are going
to construct from the Riemann curvature tensor. We will present
the vector field in the next section. Then we will discuss its
main features and the possible phenomenological consequences. A
short summary is at the end.

\section{Calculating the Fluid Flow}

To get started, it is natural to firstly seek a possible
connection between two rather different but also similar
quantities --- the vorticity of the fluid flow and the Riemann
curvature tensor. They are similar in the sense that both are
related to some non-trivial integrals along a closed path. The
vorticity is related to the exterior derivative of the fluid flow.
So to make the connection, we can also try to construct a closed
two-form from the Riemann curvature tensor. It turns out that the
time-like Killing vector also becomes a helpful ingredient in the
construction. We find that the following relation can give a
physically sensible result:
%%%
\be d\td{u}\equiv d(\rho u)=i\bpsi\gamma^{\mu\nu}\psi R_{\mu\nu
\rho\sigma} dx^\rho\wedge dx^\sigma\,,\label{def.du}\ee
%%%
where $\gamma^{\mu\nu} =-\frac{i}4 \Big[\gamma^\mu\,,\, \gamma^\nu
\Big]$. Note (\ref{def.du}) only determines $\td{u}$ up to an
exact form. This ambiguity can be fixed by imposing appropriate
boundary conditions on $\td{u}$. The spinor field $\psi$ can be
viewed as the ``square root" of the time-like Killing vector $\xi$
at the spatial infinity,
%%%
\be \bpsi\gamma^\mu\psi\rightarrow -c_\psi\xi^\mu\quad{\rm
as}\quad r\rightarrow+\infty\,,\label{psi.constr.1}\ee
%%%
where $c_\psi>0$ is a normalization constant, and $r$ is the
radial direction in a usual spherical coordinate system. For
(\ref{def.du}) to be true, one must also impose the following
constraint on the spinor field,
%%%
\be  R_{\mu\nu\rho\sigma}\pd_\alpha(\bpsi\gamma^{\mu\nu}
\psi)dx^\alpha\wedge dx^\rho\wedge dx^\sigma=0\,.
\label{psi.constr.2}\ee
%%%
In four dimensions, (\ref{psi.constr.1}) and (\ref{psi.constr.2})
contain eight equations in total. A general spinor field with
eight degrees of freedom is just enough to satisfy all the
constraints.

Note relation (\ref{def.du}) is motivated by the fact that both
$d\td{u}$ and $i\bpsi\gamma^{\mu\nu}\psi$ represent some kind of
angular momentum. The presence of the Riemann curvature tensor can
guarantee that we have a trivial result for the flat spacetime.

In the following, we will use the Kerr black hole to illustrate
what may be obtained from (\ref{def.du}). To be explicit, we write
the metric of the Kerr black hole as
%%%
\bea ds^2=\eta_{ab} e^a_{~\mu}e^b_{~\nu} dx^\mu dx^\nu
=\eta_{ab}e^ae^b\,,\quad \eta=diag\{+++-\}\,,\nn\\
e^1=f_x dr\,,\quad e^2=f_y d\theta\,,\quad e^3=f_p(d\phi-f_a
dt)\,,\quad e^4=f_t(dt-f_bd\phi)\,,\nn\\
f_x=\frac{f_y}{\sqrt{X}}\,,\quad f_y=\sqrt{r^2+a^2\cos^2\theta}\,,
\quad f_p=\frac{(r^2+a^2)\sin\theta}{f_y}\,,\quad
f_t=\frac{\sqrt{X}}{f_y}\,,\nn\\
f_a=\frac{a}{r^2+a^2}\,,\quad f_b=a\sin^2\theta\,,\quad
X=r^2+a^2-2mr\,.\label{metric}\eea
%%%
The gamma matrices are taken to be
%%%
\be\gamma^a=i\left(\begin{matrix}&\sigma^a\cr
-\sigma^a&\end{matrix}\right)\,,\;a=1,2,3\,,\quad
\gamma^4=i\left(\begin{matrix}\one_2&\cr
&-\one_2\end{matrix}\right)\,,\label{gamma.matrices}\ee
%%%
where $\sigma^{1,2,3}$ are the usual Pauli matrices and $\one_2$
is the two-dimensional unit matrix. Note the gamma matrices in
(\ref{gamma.matrices}) are in the vierbein bases. Gamma matrices
with real coordinate indices can be obtained by using
$\gamma^\mu=e_a^{~\mu}\gamma^a$ with $a=1,2,3,4$ and
$\mu=r,\theta,\phi,t$. We write the spinor field as
%%%
\bea\psi=\left(\begin{matrix} \psi_{1a}+i\psi_{1b}\cr
\psi_{2a}+i\psi_{2b}\cr \psi_{3a}+i\psi_{3b}\cr \psi_{4a}
+i\psi_{4b}\end{matrix}\right)\,,\label{ansatz.psi}\eea
%%%
where all the functions $\psi_{Ia}=\psi_{Ia}(r,\theta)$ and
$\psi_{Ib}=\psi_{Ib}(r,\theta)$ ($I=1,\cdots,4$) are real. For the
metric (\ref{metric}), we find
%%%
\bea\bpsi\gamma^r\psi&=&-\frac{2z_1}{f_x}\,,\quad \bpsi
\gamma^\phi \psi =-\frac{f_a}{1-f_af_b}\Big[\frac{2z_3}{f_a
f_p}+\frac{ z_4}{f_t} \Big]\,,\nn\\
\bpsi\gamma^\theta\psi&=&-\frac{2z_2}{f_y}\,,\quad
\bpsi\gamma^t\psi=-\frac1{1-f_af_b}\Big[\frac{2f_bz_3}{f_p}
+\frac{z_4}{f_t}\Big]\,,\eea
%%%
with
%%%
\bea z_1&=&(\phi_{2a}\phi_{3a} +\phi_{2b}\phi_{3b} +\phi_{1a}
\phi_{4a}+\phi_{1b}\phi_{4b})\,,\nn\\
z_2&=&(\phi_{2b}\phi_{3a} -\phi_{2a}\phi_{3b} -\phi_{1b}\phi_{4a}
+\phi_{1a}\phi_{4b})\,,\nn\\
z_3&=&(\phi_{1a}\phi_{3a}+\phi_{1b}\phi_{3b} -\phi_{2a}\phi_{4a}
-\phi_{2b}\phi_{4b})\,,\nn\\
z_4&=&(\phi_{1a}^2 +\phi_{1b}^2+\phi_{2a}^2 +\phi_{2b}^2
+\phi_{3a}^2 +\phi_{3b}^2+\phi_{4a}^2+\phi_{4b}^2)\,.\eea
%%%
For the Kerr metric (\ref{metric}), the time-like Killing vector
is $\xi=\pd_t$. Now (\ref{psi.constr.1}) can be satisfied with
%%%
\be z_1=z_2=z_3=0\,,\quad z_4=c_\psi(1-f_af_b)f_t\,.
\label{tmp.eq.1}\ee
%%%
The first three equations are easily solved by
%%%
\be\phi_{3b}=-\frac{\phi_{3a}}{\phi_{1b}}\phi_{1a}\,,\quad
\phi_{4a}=\frac{\phi_{3a}}{\phi_{1b}}\phi_{2b}\,,\quad
\phi_{4b}=-\frac{\phi_{3a}}{\phi_{1b}}\phi_{2a}\,.
\label{solve.psi1}\ee
%%%
From (\ref{def.du}), we find that
%%%
\be d\td{u}=f_{rh}dr\wedge d\theta +f_{pt}d\phi\wedge dt
+f_{hp}d\theta\wedge d\phi +f_{rp}dr\wedge d\phi
+f_{ht}d\theta\wedge dt +f_{rt}dr\wedge dt\,,\label{du.1}\ee
%%%
with
%%%
\bea f_{rh}&=&\frac{12amr_1\cos\theta}{f_y^4\sqrt{X}}k_1
+\frac{2mrr_3}{f_y^4\sqrt{X}}k_2\,,\nn\\
f_{pt}&=&\frac{4mrr_3\sin\theta\sqrt{X}}{f_y^6}k_1
-\frac{6amr_1\cos\theta\sin\theta\sqrt{X}}{f_y^6}k_2\,,\nn\\
f_{hp}&=&\frac{4amrr_3\sin^2\theta\sqrt{X}}{f_y^6}k_3
-\frac{8mr(r^2+a^2)r_3\sin\theta}{f_y^6}k_4\nn\\
&&-\frac{24am(r^2+a^2)r_1\cos\theta\sin\theta}{f_y^6}k_5
-\frac{12a^2mr_1\cos\theta\sin^2\theta\sqrt{X}}{f_y^6}k_6\,,\nn\\
f_{rp}&=&\frac{12am(r^2+a^2)r_1\cos\theta\sin\theta}{f_y^6
\sqrt{X}}k_3-\frac{24a^2mr_1\cos\theta\sin^2\theta}{f_y^6}k_4\nn\\
&&+\frac{8amrr_3\sin^2\theta}{f_y^6}k_5
+\frac{4mr(r^2+a^2)r_3\sin\theta}{f_y^6\sqrt{X}}k_6\,,\nn\\
f_{ht}&=&-\frac{4mrr_3\sqrt{X}}{f_y^6}k_3
+\frac{8amrr_3\sin\theta}{f_y^6}k_4
+\frac{24a^2mr_1\cos\theta\sin\theta}{f_y^6}k_5\nn\\
&&+\frac{12amr_1\cos\theta\sqrt{X}}{f_y^6}k_6\,,\nn\\
f_{rt}&=&-\frac{12a^2mr_1\cos\theta\sin\theta}{f_y^6\sqrt{X}} k_3
+\frac{24amr_1\cos\theta}{f_y^6}k_4 -\frac{8mrr_3}{f_y^6}k_5
-\frac{4amrr_3\sin\theta}{f_y^6\sqrt{X}}k_6\,,\label{def.vv}\eea
%%%
where $r_1=r^2-\frac13a^2\cos^2\theta$, $r_3=r^2
-3a^2\cos^2\theta$ and
%%%
\bea k_1&=&\phi_{1b}\phi_{3a}-\phi_{1a}\phi_{3b}
-\phi_{2b}\phi_{4a}+\phi_{2a}\phi_{4b}\,,\nn\\
k_2&=&\phi_{1a}^2+\phi_{1b}^2-\phi_{2a}^2-\phi_{2b}^2
-\phi_{3a}^2-\phi_{3b}^2+\phi_{4a}^2+\phi_{4b}^2\,,\nn\\
k_3&=&\phi_{2a}\phi_{3a}+\phi_{2b}\phi_{3b}
-\phi_{1a}\phi_{4a}-\phi_{1b}\phi_{4b}\,,\nn\\
k_4&=&\phi_{1a}\phi_{2a}+\phi_{1b}\phi_{2b}
-\phi_{3a}\phi_{4a}-\phi_{3b}\phi_{4b}\,,\nn\\
k_5&=&\phi_{2b}\phi_{3a}-\phi_{2a}\phi_{3b}
+\phi_{1b}\phi_{4a}-\phi_{1a}\phi_{4b}\,,\nn\\
k_6&=&\phi_{1b}\phi_{2a}-\phi_{1a}\phi_{2b}
-\phi_{3b}\phi_{4a}+\phi_{3a}\phi_{4b}\,.\label{def.vk}\eea
%%%
The condition $d(d\td{u})=0$ and (\ref{du.1}) lead to
%%%
\be\pd_r f_{pt}=\pd_\theta f_{pt}=0\,,\quad \pd_r f_{hp}
=\pd_\theta f_{rp}\,,\quad \pd_r f_{ht} =\pd_\theta f_{rt}\,.
\label{tmp.eq.3}\ee
%%%
For $\td{u}$ to be physically meaningful, we should in fact take
$f_{pt}=0$ in (\ref{du.1}), giving
%%%
\be \phi_{1a}^2+\phi_{1b}^2=\phi_{2a}^2+\phi_{2b}^2
=\phi_v^2\,,\quad \phi_v>0\,.\label{tmp.eq.2}\ee
%%%
The $dr\wedge d\theta$ term in (\ref{du.1}) also vanishes due to
these relations. (\ref{tmp.eq.1}) and (\ref{tmp.eq.2}) can be
solved by
%%%
\be\phi_{1a}=\sqrt{\phi_v^2-\phi_{1b}^2}\;,\quad
\phi_{2a}=\sqrt{\phi_v^2-\phi_{2b}^2}\;,\quad
\phi_{3a}=-\phi_{1b}\sqrt{\frac{c_\psi f_t}{2\phi_v^2}
(1-f_af_b)-1}\;.\label{solve.psi2}\ee
%%%
Now we are left with three undetermined functions $\phi_{1b}$,
$\phi_{2b}$, $\phi_v$ and two unsolved equations in
(\ref{tmp.eq.3}). We have achieved an extra degrees of freedom
because two equations in $d(d\td{u})=0$ are solved with the same
condition (\ref{tmp.eq.2}).

It is difficult to find exact solutions to the last two equations
in (\ref{tmp.eq.3}). It is then necessary to look for some
approximate results. In this regard, one particular interesting
case is the limit of small rotation, which should correspond to
many real life situations in nature. To study the equations in the
small rotation limit, we keep the mass $m$ fixed and take
$a\rightarrow0$. We expand the unknown functions as
%%%
\bea\phi_{1b}&=&\phi_{10}+\phi_{11}a+\phi_{12}a^2+\cdots\,,\nn\\
\phi_{2b}&=&\phi_{20}+\phi_{21}a+\phi_{22}a^2+\cdots\,,\nn\\
\phi_v&=&\phi_{v0}+\phi_{v1}a+\phi_{v2}a^2+\cdots\,.\eea
%%%
Up to the order $\cO(a^2)$, we find that
%%%
\bea\phi_{v0}&=&\frac12\sqrt{c_\psi}f^{1/4}\,,\quad
\phi_{20}=\phi_{10}\;,\quad \phi_{21}=\phi_{11}\,,\nn\\
\phi_{v1}&=&\sqrt{c_\psi}\cos\theta\frac{(33m^2+2mr
+2r^2)\sqrt{f}-2r^2}{60m^2rf^{1/4}}\,,\nn\\
\phi_{v2}&=&\sqrt{c_\psi}\Big[\frac{4m-r}{8r^3f^{3/4}}
+\frac{(1278m^5-375m^4r+140m^3r^2-120m^2r^3-8r^5)\cos^2
\theta}{3600m^4r^3f^{3/4}}\nn\\
&&\qquad+\frac{(33m^2+2mr+2r^2)\cos^2\theta}{900
m^4f^{1/4}}\Big]\,,\nn\\
\phi_{22}&=&\phi_{12}-\frac{r^2u'_{t2}}{4mf} \sqrt{c_\psi
\sqrt{f}-4\phi_{10}^2} -2\cos\theta\sin\theta\sqrt{c_\psi
\sqrt{f}-4\phi_{10}^2}\nn\\
&&\Big[\frac{2(1260m^4-180m^3r-213m^2r^2 -142mr^3-142
r^4)}{23625m^4 r\sqrt{X}}\nn\\
&&+\frac{r(9m^2+12mr+50 r^2)+24r^3\ln(1 +\sqrt{f})}{675 m^4
\sqrt{f X}}\Big]\,,\label{result.psi}\eea
%%%
where $f=1-\frac{2m}r$, and $u'_{t2}=\frac{d}{d\theta}
u_{t2}(\theta)$ is undetermined at the order $\cO(a^2)$.
Correspondingly,
%%%
\bea\td{u}&=&c_\psi(\td{u}_t dt+\td{u}_pd\phi)\,,\quad
\td{u}_\varrho=\td{u}_{\varrho0} +\td{u}_{\varrho1}a+\td{u}_{
\varrho2}a^2+\cdots\,,\quad\varrho=t,p\,,\nn\\
\td{u}_{t0}&=&\frac{4f^{3/2}}{15m}\Big(1+\frac{3m}r\Big)\,,\quad
\td{u}_{p0}=\td{u}_{t1}=\td{u}_{p2}=0\,,\nn\\
\td{u}_{p1}&=&\frac{4\sin^2\theta}{15m}\Big[1 -(1+\frac{3m}r)
f^{3/2}\Big]\,,\nn\\
\td{u}_{t2}&=&u_{t2}+\frac{8(18m^2-6mr-25r^2)\cos^2\theta}{675
m^3r^2}\nn\\
&&+\frac{2\sqrt{f}(20m^4+5m^3r+3m^2r^2+2mr^3+2r^4)}{45m^3r^4}\nn\\
&&+\frac{4\cos^2\theta\sqrt{f}(6930m^4-1215m^3r-1149m^2r^2
+284mr^3+284r^4)}{23625m^3r^4}\nn\\
&&-\frac{32\cos^2\theta}{225m^3}\ln(1+\sqrt{f})
\,.\label{result.u.1}\eea
%%%
In the limit $a\rightarrow0$, the black hole horizon is located at
$r_0\approx2m-\frac{a^2}{2m}$. The result in (\ref{result.psi})
diverges at $f=0$, i.e., $r=2m$. We expect such a divergence to
occur exactly on the horizon for the exact solution. As we will
explain in the next section, such divergence is not a problem for
us because we can only trust our result outside the black hole
horizon.

Another more fatal divergence may come in the form $\ln r$, as is
hinted by the presence of $\ln(1+\sqrt{f})$ in (\ref{result.psi})
and (\ref{result.u.1}). This will indicate that the result is not
well defined at the spatial infinity. We have checked that such
divergence can be removed at the order $\cO(a^3)$, by
appropriately adjusting the function $u_{t2}$. But it is not
certain if the required cancellation exists to all orders in the
expansion of $a$. It is then very important if one can solve the
last two equations in (\ref{tmp.eq.3}) exactly. In the following,
we will assume that such divergence can always be removed.

In any case, it is already quite remarkable that, as we will show
in the next section, the leading order results from (\ref{def.du})
does display features that look like a realistic physical
quantity. This is why we think that we might be on the right
track. We will discuss the result in more detail in the next
section.

\section{Features of the Fluid Flow}

Our most important quantity is the fluid flow. From (\ref{def.du})
and (\ref{result.u.1}), we find that
%%%
\bea\rho&=&\frac{4c_\psi}{15m}f\Big(1+\frac{3m}r\Big) +\cO(a^2)\,,\nn\\
u^t&=&\frac1{\sqrt{f}}+\cO(a^2)\,,\quad u^r=u^\theta=0\,,\nn\\
u^\phi&=&\frac{a}{r^2 \sqrt{f}} \Big[1-f^{-1/2}(1+\frac{3m}r)^{-1}
\Big] +\cO(a^2) \,.\label{fluid}\eea
%%%
Now let's look at the most important features of the result:
\begin{itemize}
\item Firstly, $u\rightarrow\pd_t$ as $r\rightarrow +\infty$. This
means that the fluid is static at the spatial infinity. This is
exactly what we expect for the fluid underlying our spacetime.
What's more, one also has $u^\phi\sim a$, which means that there
is no spatial flow underlying the Schwarzschild black hole. This
also fits very well with our intuitive picture about the fluid.
These two points are the most important reasons making us think
that (\ref{fluid}) might be describing something physical.

\item Secondly, $\rho\rightarrow0$ as $f\rightarrow0$. For the
exact solution, we expect the density to vanish exactly on the
horizon. If we take the fluid picture seriously, then this result
indicates that the geometrical picture is not valid on the
horizon. The reason is the following:

If the spacetime is emergent from some underlying condensed matter
system, then the geometrical picture is only valid in the large
distance limit, where the discrete nature of the underlying
material is irrelevant. Now if $\rho\rightarrow0$ at some point in
the emergent spacetime, it does not mean that the density of the
fluid really vanishes there. It only means that we have come to a
location where a tiny region of the underlying material is
projected to a huge region in the emergent spacetime. To be more
explicit, Let's use the Schwarzschild black hole as an example. It
is reasonable to assume that the underlying material is best
described by a flat metric, which coincides with that of the
Schwarzschild black hole at the spatial infinity. Let's also
assume that the density of the underlying material is everywhere
finite in the flat metric. Near the black hole horizon, a shell of
depth $dr$ of the underlying material is mapped to a shell of
depth $\frac{dr}{\sqrt{f}}$ in the Schwarzschild spacetime. The
density of the underlying material thus vanishes on the horizon
when ``observed" from the Schwarzschild spacetime. If this picture
is true, then we are probing smaller and smaller distances in the
underlying material as we come closer and closer to the black hole
horizon. The discrete nature of the underlying material will
become important and the geometric picture will break down before
the horizon is reached. For this reason, we will only trust our
result outside the black hole horizon.

\item Thirdly, $u^\phi$ falls as $r^{-2}$ when $r\rightarrow
+\infty$, but it diverges at $f=0$. Similarly, $u^t$ also diverges
at $f=0$. For the reason explained above, the divergence on the
horizon should not be a big concern for us, because we can only
trust our result out side the black hole horizon.

\item Finally, $\rho\rightarrow4c_\psi/15m$ as
$r\rightarrow+\infty$ . At the spatial infinity, the density of
the fluid is that of a flat spacetime, and should not depend on
any physics near the center. As a result, we expect $c_\psi\propto
m$. In a more general case, we expect $c_\psi$ to be proportional
to the total energy contained in the geometry.
\end{itemize}
Now we see that (\ref{fluid}) fulfills all our expectations about
the speculated fluid flow, and we also have a consistent
explanation to all the displayed features.

To this end, it is also interesting to note that if
%%%
\be \zeta^\mu=-\bpsi\gamma^\mu\psi\,,\label{def.zeta1}\ee
%%%
then (\ref{solve.psi1}) and (\ref{solve.psi2}) lead to
%%%
\be \zeta=c_\psi(f_a\pd_\phi+\pd_t)\,.\ee
%%%
We see that $c_\psi^{-1}\zeta$ interpolates between the time-like
Killing vector at the spatial infinity and the null Killing vector
on the horizon. Most interestingly, if we let
%%%
\be \zeta^\mu=\rho u^\mu\,,\label{def.zeta2}\ee
%%%
then
%%%
\bea\rho&=&c_\psi\frac{f_y\sqrt{X}}{r^2+a^2}\,,\quad
u=\frac{c_\psi}{\rho}(f_a\pd_\phi+\pd_t)\,. \label{fluid2}\eea
%%%
Just like (\ref{fluid}), this result also displays all the
interesting features listed below (\ref{fluid}). Thus,
(\ref{fluid2}) is another possible candidate description of the
speculated fluid flow.

One potential advantage of (\ref{def.zeta1}) and (\ref{def.zeta2})
over (\ref{def.du}) is that they are easier to generalize to
higher dimensions. For example, for the Myers-Perry black hole in
five dimensions \cite{myers.perry86},
%%%
\bea ds^2=\eta_{ab} e^a_{~\mu}e^b_{~\nu} dx^\mu dx^\nu
=\eta_{ab}e^ae^b\,,\quad \eta=diag\{++++-\}\,,\nn\\
e^1=\frac{\sqrt{r^2+y^2}}{\sqrt{X}}dr\,,\quad
e^2=\sqrt{r^2+y^2}d\theta\,,\nn\\
e^3=\frac{\cos\theta\sin\theta}{y\sqrt{r^2+y^2}} \Big[\frac{a^2
(d\phi_1-f_a dt)}{f_a} -\frac{b^2(d\phi_2 -f_b dt)}{f_b}\Big]\,,\nn\\
e^4=\frac{a b}{r y}\Big[\frac{\sin^2\theta(d\phi_1-f_a dt)}{f_a}
+\frac{\cos^2\theta(d\phi_2 -f_b dt)}{f_b}\Big]\,,\nn\\
e^5=\frac{\sqrt{X}}{\sqrt{r^2+y^2}}\Big[dt-a\sin^2\theta d\phi_1
-b\cos^2\theta d\phi_2\Big]\,,\nn\\
f_a=\frac{a}{r^2+a^2}\,,\quad f_b=\frac{b}{r^2+b^2}\,,\quad
X=\frac{(r^2+a^2)(r^2+b^2)}{r^2}-2m\,,\nn\\
y=\sqrt{a^2\cos^2\theta +b^2\sin^2\theta}\,,\label{metric2}\eea
%%%
one can check that (\ref{def.zeta1}) leads to (with some simple
choice of the spinor degrees of freedom)
%%%
\be\zeta=c_\psi(f_a\pd_{\phi_1}+f_b\pd_{\phi_2}+\pd_t)\,.\ee
%%%
Again, $c_\psi^{-1}\zeta$ interpolates between the time-like
Killing vector at the spatial infinity and the null Killing vector
on the horizon. Using (\ref{def.zeta2}), we can also interpret the
result as a fluid flow underlying the five dimensional Myers-Perry
black hole.

\section{Summary}

In this note, we consider an emergent picture of the spacetime:
the material underlying our spacetime is assumed to be a condensed
matter system (a fluid), all known matter fields are excitations
in the fluid, and our observed spacetime is derived from the
equations obeyed by the matter fields.

We are particularly interested in possible connections between the
geometrical properties of the emergent spacetime and the
properties of the underlying fluid. We have presented some partial
results in support of such possible connections. In particular, we
have constructed from the Riemann curvature tensor a vector field,
which behaves just like the speculated fluid flow. Now if we take
the fluid picture seriously, then our result indicates that the
geometric picture breaks down on the horizon, and we can only
trust our result outside the black hole horizon.

In the calculation, we have also noticed a second possible
formulation of the fluid flow. It also displays all the desired
quantitative features, and is more easily generalized to other
dimensions of the spacetime. We shall consider this formulation in
more detail in future works.

\section*{Acknowledgement}

This work was supported by the Alexander von Humboldt-Foundation.

%\newpage

%%%%%%%%%%%%%%%%%%%%%%%%%%%%%%%%%%%
\end{document}